\documentclass[reprint,twocolumn,pra,floatfix,superscriptaddress]{revtex4-1}
\usepackage{multirow}
\usepackage[latin9]{inputenc}
\usepackage{color}
\usepackage{amssymb,amsmath,amscd,amsthm,amsfonts,}
\usepackage{amstext}
\usepackage{graphicx}
\usepackage[table]{xcolor}
\usepackage{graphics}
\usepackage{epsfig}

\makeatletter
\@ifundefined{textcolor}{}
{%
 \definecolor{BLACK}{gray}{0}
 \definecolor{WHITE}{gray}{1}
 \definecolor{RED}{rgb}{1,0,0}
 \definecolor{GREEN}{rgb}{0,1,0}
 \definecolor{BLUE}{rgb}{0,0,1}
 \definecolor{CYAN}{cmyk}{1,0,0,0}
 \definecolor{MAGENTA}{cmyk}{0,1,0,0}
 \definecolor{YELLOW}{cmyk}{0,0,1,0}
}

\newtheorem{theorem}{Theorem}[section]

\newcommand{\PT}{\mathcal{PT}}

\newcommand{\tz}{\tilde{z}}

\newcommand{\rev}[1]{\textcolor{black}{#1}}

\makeatother

\begin{document}

\title{
 {Designing lasing and perfectly absorbing potentials}
}

\author{Vladimir V. Konotop}
\email[]{vvkonotop@fc.ul.pt}
\affiliation{
	Departamento de F\'{i}sica and Centro de F\'{i}sica Te\'orica e Computacional, Faculdade de Ci\^encias, Universidade
	de Lisboa, Campo Grande, Ed. C8, Lisboa 1749-016, Portugal
	}
\author{Evgeny Lakshtanov}
\email[]{lakshtanov@ua.pt}
\affiliation{
	CIDMA, Department of Mathematics, University of Aveiro, Aveiro 3810-193, Portugal and MathLogic LTD, London, UK.
}
\author{Boris Vainberg}
\email[]{brvainbe@uncc.edu}
\affiliation{Department of Mathematics and Statistics, University of North Carolina,
Charlotte, NC 28223, USA.
}

\date{\today}


\begin{abstract}

Existence of a spectral singularity (SS) in the spectrum of {a Schr\"{o}dinger operator with} a non-Hermitian potential requires exact matching of parameters of the potential. 
We provide a necessary and sufficient condition for a potential to have a SS at a given wavelength. It is shown that potentials with SSs at prescribed wavelengths can be obtained by a simple and effective procedure. In particular, the developed approach allows one to obtain potentials with several SSs and with SSs of the second order, as well as potentials obeying a given symmetry, say, $\PT-$symmetric potentials. Also, the problem can be solved when it is required to obtain a potential obeying a given symmetry, say, $\PT-$symmetric potential. We illustrate all these opportunities with examples. We also describe splitting  of a second-order SSs under change of the potential parameters, and discuss possibilities of experimental observation of SSs of different orders.
	 
\end{abstract}

\maketitle

\section{Introduction}
\label{sec:introdcution}

Spectral singularities (SSs) are characteristic features of  {continuous spectra} { of Schr\"odinger operators with} complex potentials. They were  discovered more than six decades ago~\cite{Naimark},  {coined as such in~\cite{SS_mathematics}}, and studied in mathematics since then~\cite{SS_mathematics,68,Guseinov}. Although the first examples of physical systems with singularities of scattering data,  {i.e., with SSs,} were also reported a few decades ago~\cite{first}, the relevance of SSs to physical application became widely recognized only recently. It was argued that SSs are related to lasing~\cite{lasing,Longhi,Stone2011} and to the coherent perfect absorption, which can be interpreted as time-reversed lasing,~\cite{StonePRL,StoneScience} (see  \cite{review} for a recent review). 

\rev{Mathematically, a {\em SS is a real pole $k_0\in\mathbb{R}$ of the truncated (see Sec.~\ref{sec:general}) resolvent $(H-k^2I)^{-1}$ of a Hamiltonian $H$}, where $I$ is the identity operator. In the one-dimensional (1D) case,  SSs can be characterized as zeros of the $M_{22}(k)$-element of the transfer matrix $M(k)$ of a non-Hermitian potential. For discussion of different aspects of the definition of a SS see also~\cite{2delta}, as well as Sec.~\ref{sec:general} below.  Importantly, {\em a SS belongs to a continuous spectrum}. It must be distinguished from the notion of an exceptional point (EP), which was introduced in~\cite{Kato} approximately at the same time as SSs. {\em An EP is a point of the discrete spectrum} where two (or more) eigenfunctions coalesce in a degeneracy point making the geometric degeneracy of the spectral eigenvalue smaller than its algebraic  one.}   EPs also received significant attention in physical literature~\cite{EP}. Consideration of potentials possessing EPs, however, goes beyond the scope of the present paper.
	 
Thus, the problem of constructing a non-Hermitian potential which coherently absorbs or emits radiation, is reduced to designing a potential which possesses a prescribed SS. Note, that similar types of inverse problems, i.e. finding potentials featuring specific properties, was explored in different physical applications. In particular, construction of non-Hermitian potentials possessing entirely linear spectrum was explored in~\cite{SUSY} while complex potentials supporting nonlinear waves were addressed in~\cite{AKSY}.

Existence of singularities in a spectrum of an operator with a complex potential is a delicate property: not any complex potential features it. Usually finding such singularities require numerical calculation of the spectrum. Some examples, however are known explicitly. In particular, these are potentials in a form of Dirac deltas with complex amplitude~\cite{DiracDelta,2delta} in one dimension, as well as their multidimensional extensions~\cite{KonZez18}, rectangular potentials~\cite{rectang}, and the complex Scarf II potential~\cite{ScarfII}. The most of the explored examples feature only one SS. In particular, the existence of at most one SS in parity-time ($\PT$) symmetric potentials was conjectured recently~\cite{PT-sing}. Potentials with two spectral singularities have been reported too~\cite{ScarfII-two-singul}. 

In terms of physical applications, a SS is characterized by a specific wavelength (wavenumber or  frequency) at which the scattering data display singularities. The goal of the present paper is to show that for a SS with an {\it a priori} given wavelength, multi-parametric families of complex potentials of a quite general type can be constructed. Furthermore, we show how it is possible to obtain families of tunable potentials featuring two (or more) spectral singularities at desirable wavelengths, as well as potentials with SSs of the second order. We also discuss how SS of the second order   split when the system parameters are perturbed. Our main attention is focused on 1D scattering problems, although some general statements will be formulated, too.    
 
The organization of the paper is as follows. In Sec.~\ref{sec:general} we formulate necessary and sufficient conditions for a scattering problem to have a SS. In Sec.~\ref{sec:1Dexamples} we focus on obtaining 1D potentials with prescribed SSs. This section includes examples of the known cases, like a rectangular potential, as well as more general even and $\PT-$symmetric potentials. In Sec.~\ref{sec :pot_SS_second_order} we describe a general approach for obtaining potentials with SSs of the second order. In Sec.~\ref{sec:twodelta} using the known example of a potential in a form of two Dirac delta-functions, we discuss splitting of SSs of the second order. Scattering of wavepackets by SSs of different orders, allowing one to distinguish an order of a SS, is described in Sec.~\ref{sec:double_SS_detect}. The outcomes are summarized in Conclusion. 

\section{General consideration}
\label{sec:general}

Starting with a $d$-dimensional case, $x\in \mathbb{R}^d$,  we consider the operator
\begin{eqnarray}
\label{eq:H}
H=-\Delta + U(x),  
\end{eqnarray}
where $\Delta$ is the $d$-dimensional Laplacian and $U(x)$ is a bounded complex potential decaying exponentially, or faster, at infinity.  The spectrum of $H$ consists of the continuous part $[0,\infty)$ and a finite set of eigenvalues $\{\lambda_j\}$, see \cite[Lemma 8 and Theorem 10]{75}. Due to the Kato theorem~\cite{KatoTheorem} (which is valid for  potentials $U$ decaying fast enough at infinity), $\lambda_j\notin (0,\infty)$. In the case of a compactly supported potential $U$, i.e., of a potential that vanishes outside of a given bounded domain in $\mathbb{R}^d$,  a simple proof of the Kato theorem  can be found in \cite[Theorem 3.3]{66}. 
The resolvent $R_\lambda=(H-\lambda I)^{-1}:L^2(\mathbb{R}^d)\to L^2(\mathbb{R}^d)$, where $I$ is the identity operator and $\lambda$ is the spectral parameter, of the operator $H$ is meromorphic in $\lambda\notin [0,\infty)$ with poles at $\lambda=\lambda_j$. Hence, for $\lambda=k^2$, the resolvent $R_{k^2}$ is meromorphic in $k$ in the upper half plane Im$k>0$. 

Let us define the truncated resolvent $\widehat{R}_\lambda=\alpha(x)R_\lambda\alpha(x)$, where $\alpha(x)$ is a characteristic function of a ball $B_\rho=\{x:|x|<\rho\}$, i.e., $\alpha(x)=1$ in $B_\rho, ~\alpha(x)=0$ outside of $B_\rho$. The idea of this truncation is in the cutting off the behavior of the functions at infinity: if the support of $f$ belongs to $ B_\rho$, then $u=\widehat{R}_\lambda f$ is the solution of $(H-\lambda)u=f$ from $L^2(R^d)$ restricted to the ball $B_\rho$.
The truncated resolvent admits a meromorphic continuation in $\lambda$ through the continuous spectrum on another sheet of the Riemannian surface. To be more exact, $\widehat{R}_{k^2}$ is meromorphic in the complex $k$-plan when the dimension $d$ is odd, and it is meromorphic on the Riemannian surface of $\ln k$ if $d$ is even, see \cite[Theorems 3, 4]{68}. 
 Poles of $\widehat{R}_\lambda $ can serve as a rigorous definition (and a key to a study) of SSs.
The symmetry arguments and the Kato theorem imply that $\widehat{R}_{k^2}$ does not have real positive poles $k\neq 0$ if $U$ is real valued. Hence there are no SSs for real fast decaying potentials. A simple example of a complex valued potential $U$ for which $\widehat{R}_{k^2}$ has a pole at $k=k_0\neq 0$ can be found in \cite[Chapter 3, \%3]{66}. 

A starting point for constructions below is the following statement, proved in \cite[Theorems 3, 4]{68}:
\begin{theorem} \label{t0} 
Operator $\widehat{R}_{k^2}$ has a pole at real $k=k_0\neq 0$ if and only if the problem  
\begin{subequations}
\begin{eqnarray}
\label{1}
-\Delta \psi+ U(x)\psi=k_0^2\psi,\quad x\in \mathbb{R}^d,  
\\ 
\psi=\frac{f(x/r)}{r^{(d-1)/2}}e^{ik_0r}+O\left(\frac{1}{r^{(d+1)/2}}\right), \nonumber \\ r=|x|\to\infty, \label{1aa} 
\end{eqnarray}
\end{subequations}
 where the amplitude $f(x)\in C^\infty(S),~ S$ is the unit sphere in $d$-dimensional space, has a non-trivial solution.
 \end{theorem}
Formally, the function $\psi(x)$ describes a coherent perfect absorber (CPA) solution (without reflected and transmitted waves) if $k_0<0$  and lasing solution (without waves incident on the potential) if $k_0>0$. 

Below, we are particularly interested in less common situations when a SS is of the second order. Obtaining respective potentials will be based on the theorem, similar to Theorem \ref{t0}, whose proof, however, is more involved.
\begin{theorem} \label{tt}
	Operator $\widehat{R}_{k^2}$ has a pole of order $m\geq 2$ at real $k=k_0\neq  0$ if and only if  the following problem has a non-trivial solution
	\begin{subequations}
		\begin{eqnarray}\label{2}
	-\Delta \phi+ U(x)\phi-k_0^2\phi=2k_0\psi,\quad x\in \mathbb{R}^d, 
	\\  
\label{2aa}
\phi=\frac{if( x/r)}{r^{(d-3)/2}}e^{ik_0r}+\frac{g(x/r)}{r^{(d+1)/2}}e^{ik_0r}
\nonumber \\ 
+O\left(\frac{1}{r^{(d-1)/2}}\right),
	\end{eqnarray}
	\end{subequations}
	where $\psi$ is a solution of (\ref{1}),(\ref{1aa}), $f$ is its amplitude  and $g\in C^\infty(S)$. 
\end{theorem}
Note that $\phi$ decays at infinity slower than $\psi$ does.

\section{One-dimensional linear problems}
\label{sec:1Dexamples}

It is particularly simple to apply the above statements to 1D problems. When $x\in\mathbb{R}$ and $U$  has a compact support, without loss of generality, one can consider the potential to be localized in the interval $(-1,1)$. Then problem (\ref{1}), (\ref{1aa}) takes the form
\begin{subequations}
	\label{eq:u0}
\begin{eqnarray}
\label{1a}
 &&-\psi''+U(x)\psi=k_0^2\psi, \quad |x|<1 ,
 \\
 \label{1b}
 && \psi=\alpha_\pm e^ {ik_0 |x| }, \quad \pm x \geq 1
\end{eqnarray}
\end{subequations}
(hereafter a prime stands for the derivative with respect to $x$). 
Now, the operator $H=-\frac{d^2}{dx^2}+U(x)$ has a pole of the truncated resolvent $\widehat{R}_{k^2}$ at real $k=k_0\neq 0,$ if and only if  (\ref{eq:u0})  has a non trivial solution. Problem (\ref{eq:u0}) on the whole line can be reduced to the Sturm-Liouville problem on the interval $[-1,1]$:
\begin{subequations}
	\label{eq:z}
\begin{eqnarray}
\label{eq:z1}
&&-z''+U(x)z=k_0^2z, \quad x\in (-1,1); 
\\ 
\label{eq:z2}
&&z'\mp ik_0 z=0, \quad  \mbox{at $x=\pm 1$}.
\end{eqnarray}
\end{subequations}
If such a $z$ exists, then the solution $\psi$ of (\ref{eq:u0}) can be defined as follows
\begin{eqnarray}
\label{eq:solut}
\psi(x)=\left\{
\begin{array}{ll}
z(x) & |x|\leq 1, 
\\[1mm]
z(\pm 1)e^{ik_0(|x|-1)}, & \pm x \geq 1. 
\end{array}
\right.
\end{eqnarray}
Similarly, one dimensional version of (\ref{2}), (\ref{2aa}) has the following form:
\begin{subequations}
\label{v12}
\begin{eqnarray}
\label{v1}
&&
-\phi''+U(x)\phi-k_0^2\phi=2k_0\psi, \quad |x|< 1,
\\ 
\label{v1b}
&& \phi =\left(i\alpha_{\pm}|x|+\beta_{\pm}\right)e^{ik_0|x|}, \quad \pm x\geq 1.
\end{eqnarray}
\end{subequations}

In order to relate the above results to the scattering characteristics of the potential $U(x)$, and more specifically to its transfer matrix $M(k)$, one considers a solution $\psi=\psi(x;k)$ of the scattering problem
\begin{subequations}
\label{eq:1Dscat}
\begin{eqnarray}
\label{eq:1Dscat1}
 && -\psi''+U(x)\psi=k^2\psi, \quad |x|<1
\\ 
\label{eq:1Dscat2}
&& \psi=a_\pm e^ {ik x}+b_\pm e^ {-ik x}, \quad \pm x \geq 1.
\end{eqnarray}
\end{subequations}
and uses the standard definition of the transfer matrix through the relation:
\begin{eqnarray}
\label{transfer}
\left( 
\begin{array}{c}
a_+(k)\\b_+(k)
\end{array}\right)=M(k)\left( 
\begin{array}{c}
a_-(k)\\b_-(k)
\end{array}\right).
\end{eqnarray}
Problem (\ref{eq:u0}) has a non-trivial solution for real $k_0\neq 0$ if and only if $M_{22}(k_0)=0$. The latter is equivalent to  $M_{11}(-k_0)=0$. 
Usually laser wavelengths, $\lambda_0=2\pi/|k_0|$, are determined by zeroes $M_{22}(k_0)=0$ at $k_0>0$, while CPA frequencies are determined from $M_{11}(|k_0|)=0$ at $k_0<0$.

Now we can formulate an algorithm of constructing a complex potential $U(x)$, having a SS with {\em a priori} given properties. To this end,  one can start with $\psi(x)$ of the form (\ref{eq:solut}) where $z\in C^2([-1,1])$ is an arbitrary function that satisfies conditions (\ref{eq:z2}) and does not vanish on $[-1,1]$ [the latter condition can be weaken: it is enough to know that the left hand side of Eq.~(\ref{5}) below is bounded]. Obviously, the set of functions $z(x)$ with described properties is very wide. Then, one  substitutes the chosen $z(x)$ into (\ref{eq:z1}) and defines $U(x)$ from the formula:
\begin{eqnarray}\label{5}
U(x)=\frac{z''+k_0^2z}{z},~~|x|<1.
\end{eqnarray}
The obtained potential $U(x)$ has a SS at $k=k_0$.

\subsection{Even potentials with simple spectral singularities}
\label{sec:simple_SS}

 \subsubsection{Rectangular potentials.}
 \label{subsec:rectangular}
 
  Turning to examples, we start  by recovering the known results for a rectangular potential~\cite{rectang}, however with the difference that our starting point will be the wavevector $k_0$, at which the laser or CPA must occur. To this end we choose $z=e^{i\kappa x}+Ce^{-i\kappa x}$ with a complex $\kappa$ and obtain the complex potential $U_{\rm rect}=k_0^2-\kappa^2$ at $|x|\leq 1 $ and $U_{\rm rect}=0$ at $|x|> 1$, provided $\kappa$ is obtained from one of the transcendental equations
\begin{eqnarray}
\label{rect_a}
i\kappa\tan\kappa=k_0, \quad  -i\kappa\cot\kappa=k_0
\end{eqnarray}
and $C=e^{i\kappa x}(\kappa-k_0)/(\kappa+k_0)$. Since each of Eqs. (\ref{rect_a}) has infinite number of solutions, there exist  infinite number of rectangular potentials possessing SS at $k_0$.

 \subsubsection{Varying even potentials.} 
 \label{subsec:varying}
 
 Now we consider more general examples, starting with families of potentials depending on two arbitrary functions, but having a fixed SS $k_0$. To this end we look for a function $z$ satisfying (\ref{eq:z2}) in the form
 \begin{eqnarray}
 \label{aux1}
 z(x)=\gamma(x)\cos (k_0x)+i\nu(x) \sin (k_0 x)
 \end{eqnarray}
 where $\gamma(x)$ and $\nu(x)$ are real functions, which are arbitrary, so far. Imposing the boundary conditions at $x=\pm 1$ we arrive at the following set of constraints
 \begin{subequations}
 \label{constraints}
   \begin{eqnarray}
 \gamma'(1)\cos k_0 &=& [\gamma(1)-\nu(1)] k_0\sin k_0,
 \\ 
 \nu'(1)\sin k_0&=&[\gamma(1)-\nu(1)] k_0\cos k_0.
 \\
 \gamma'(-1)\cos k_0&=&-[\gamma(-1)+\nu(-1)] k_0\sin k_0,
 \\ 
 \nu'(-1)\sin k_0&=&[\gamma(-1)+\nu(-1)] k_0\cos k_0.
 \end{eqnarray}
  \end{subequations}
We note that these conditions do not depend on the sign of $k_0$ i.e. they are the same for laser and CPA solutions.

\begin{figure}
	\includegraphics[width=0.49\columnwidth]{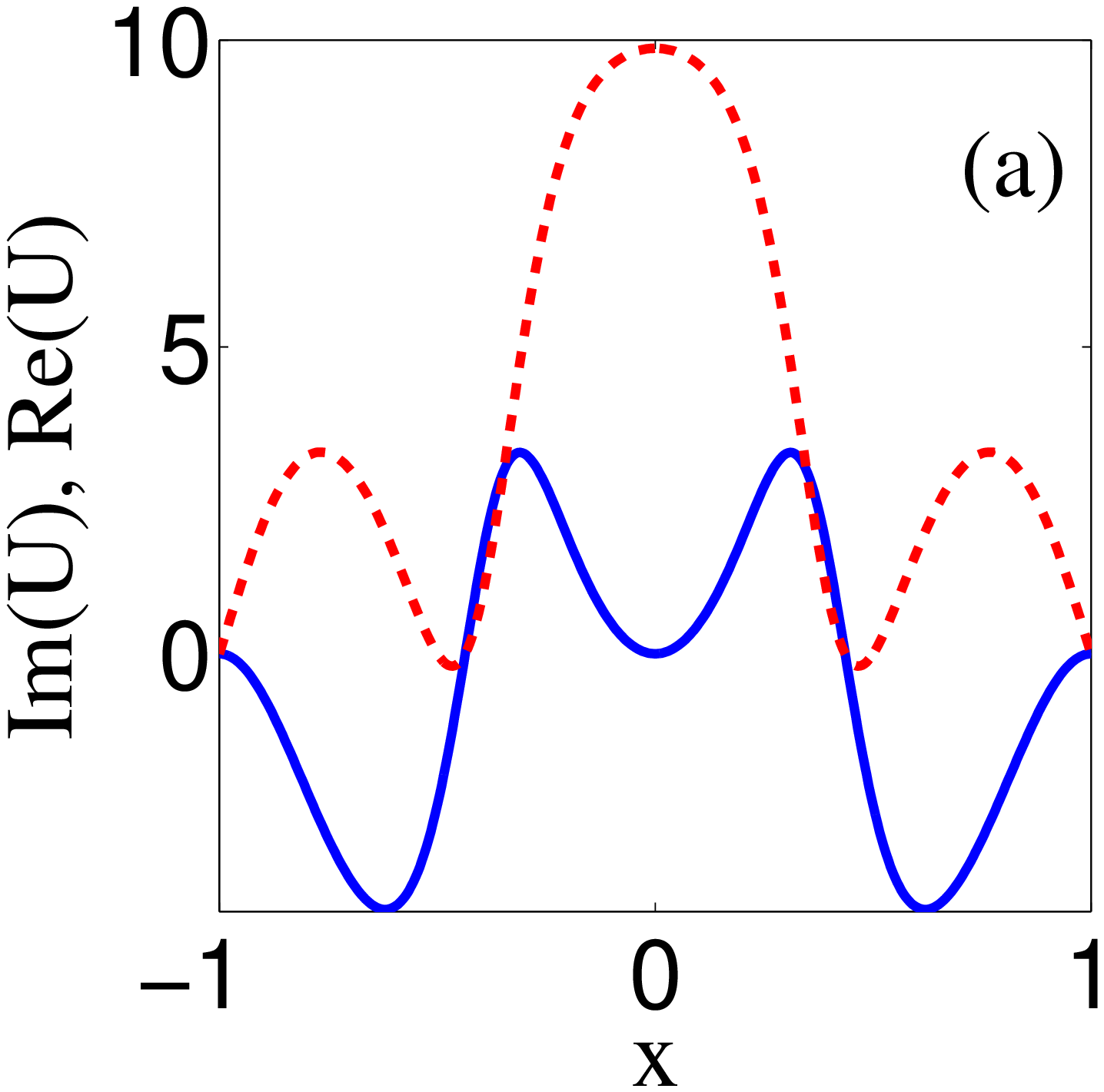}
	\includegraphics[width=0.49\columnwidth]{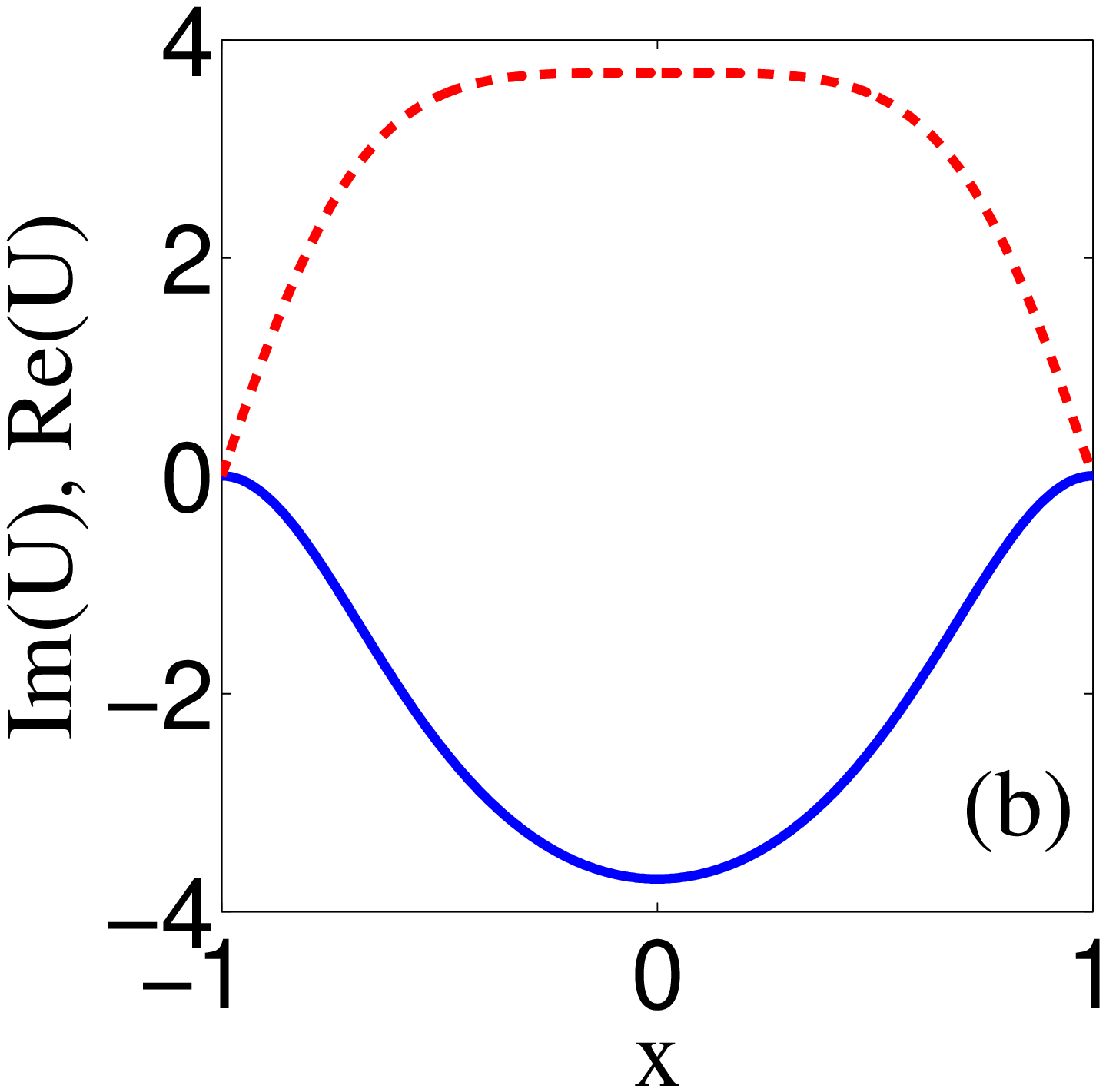}
	\\
	\includegraphics[width=0.49\columnwidth]{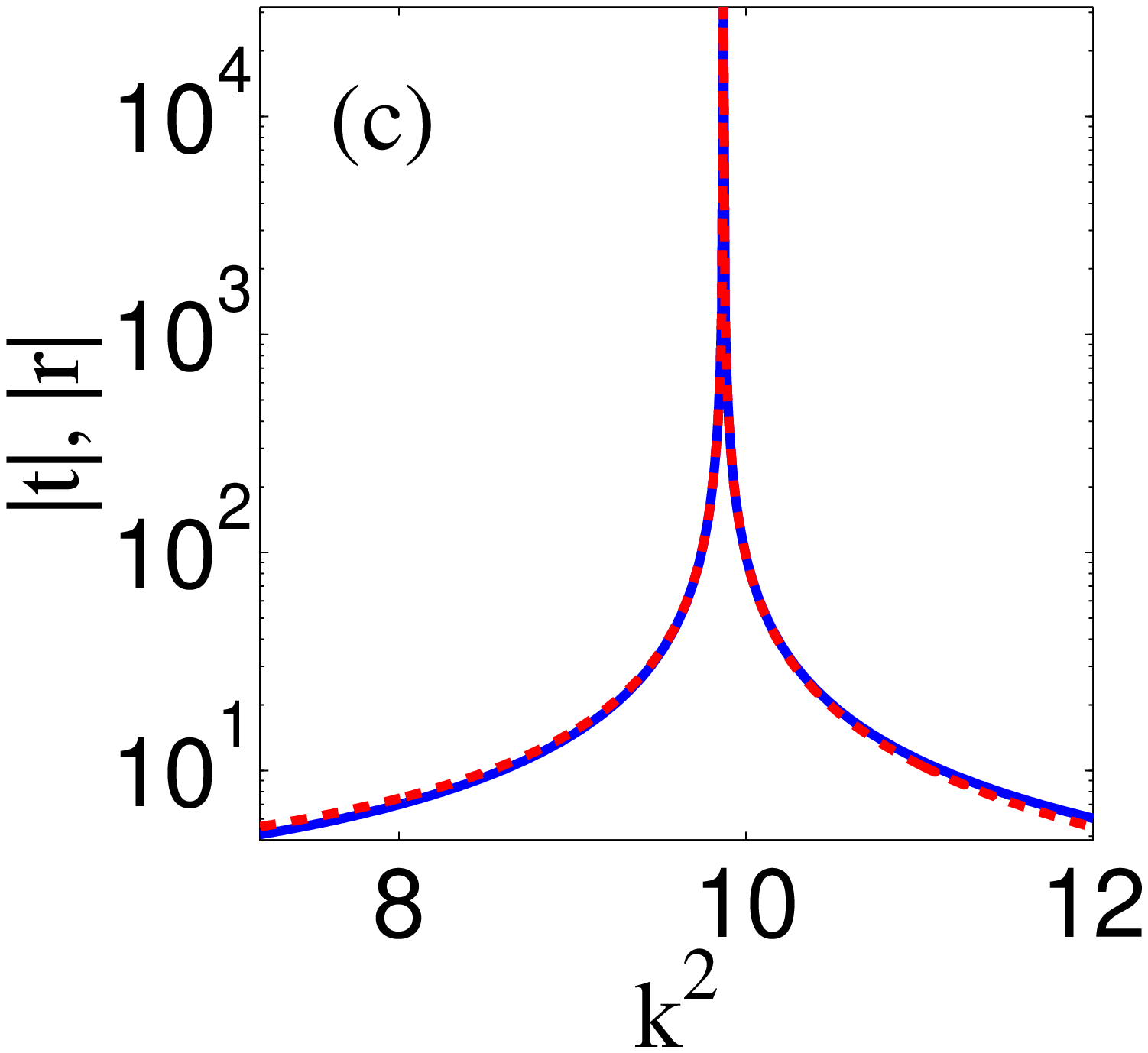}
	\includegraphics[width=0.49\columnwidth]{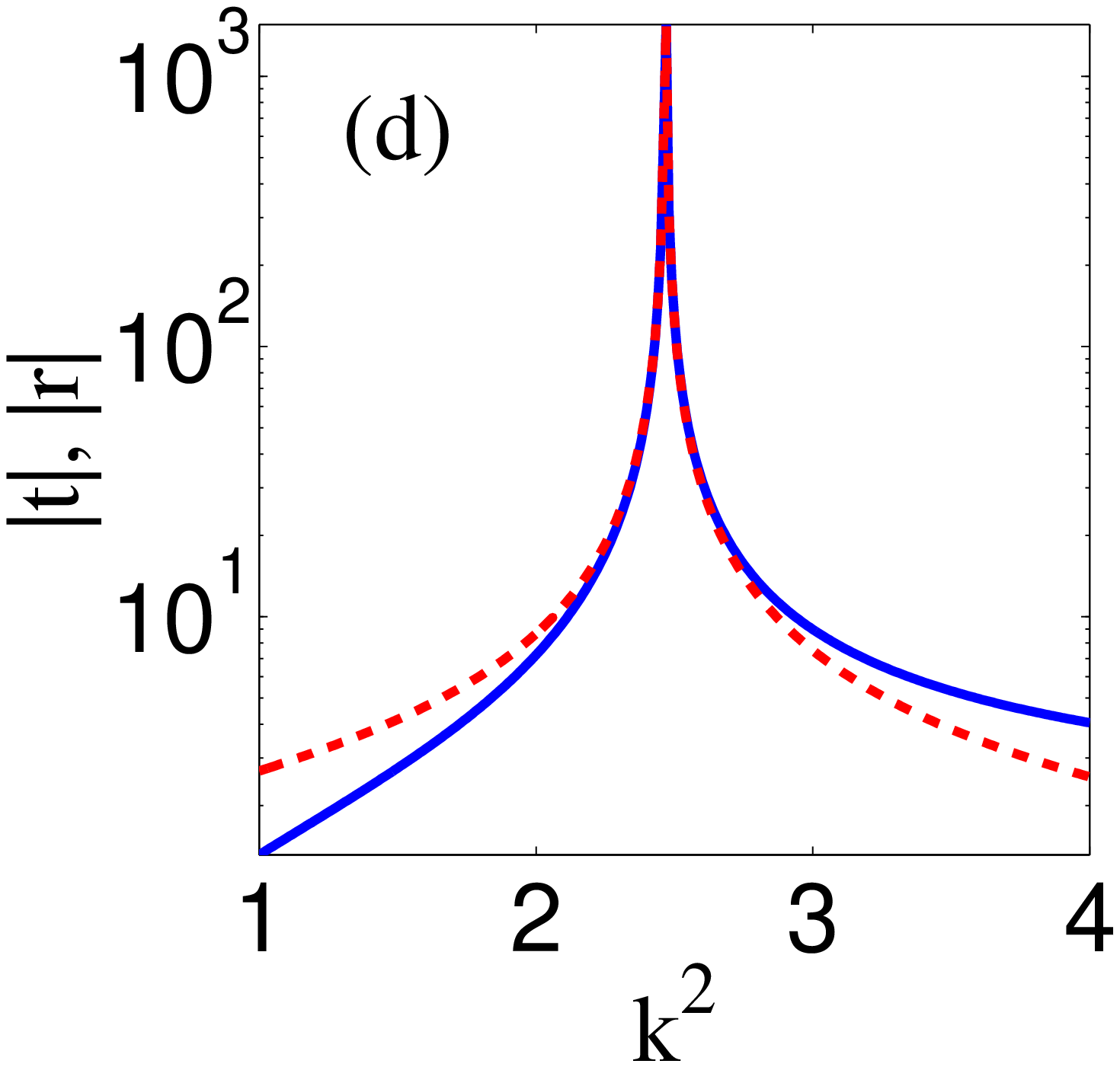}
	\caption{Even non-Hermitian potential (\ref{pot1}) having SS at $k_0=\pi$ (a) and potential (\ref{pot2}) having SS at $k_0=\pi/2$ (b). The real and imaginary parts of the potentials are shown by blue solid and red dashed lines, respectively. \rev{(c) and (d) the absolute values of the transmission coefficient $|t|$ (solid blue lines) and of the reflection coefficients $|r_{\rm L}|=|r_{\rm R}|$ (dashed red lines) for the potentials shown in (a) and (b), respectively.} }
	\label{fig:one}
\end{figure}

The simplest potentials, generated by the functions $z(x)$ of type (\ref{aux1}), are found particularly easily, if the resonant condition $k_0=\pi n$ for either emitted ($n$ is positive integer)  or coherently absorbed ($n$ is negative integer) radiation, is satisfied. Note, that in the physical units this requirement corresponds to the potential size of $|n|$ wavelengths: $L=\lambda_0 |n|$. In that case the potential providing the required SS is obtained using
\begin{equation}
\label{eq:ansatz1}
z= \cos(\pi n x)+i\sin\left(\frac{\pi }{2}x\right) \sin (\pi n x)
\end{equation}
in (\ref{5}). For the simplest case of $n=1$ we obtain:
\begin{equation}
\label{pot1}
U=\frac{{\pi}^{2}}{4}\frac {  3\cos \left( \pi x/2 \right) +5
		\cos \left( {3\pi x}/{2} \right)   }{\cos \left( \pi x/2
		\right) -\cos \left( 3\pi x/2 \right) -2i\cos \left( \pi x
		\right) }.
\end{equation}
If the potential is chosen to have the size measured in half-wavelengths, i.e., $k_0=\pi\left(\frac 12 +n\right)$, what in the physical units means  $L=\lambda_0 \left(\frac 12 +n\right)$, then the function $z$ can be chosen as  
\begin{equation}
\label{eq:ansatz2}
z= \sin\left(\frac{\pi }{2}x\right)\cos \left[ \left(\frac 12 +n\right)\pi x\right]
+i  \sin\left[ \left(\frac 12 +n\right)\pi x\right]
\end{equation}
(here $n$ is any integer including $n=0$). In the simplest case of $n=0$, the   potential has the form
\begin{eqnarray}
\label{pot2}
U=-\frac {3{\pi}^{2}}{4}\frac {\cos \left( \pi x/2 \right) }{\cos \left( 
		\pi x/2 \right) +i}.
\end{eqnarray}
The real and imaginary parts of potentials (\ref{pot1}) and (\ref{pot2}) are shown in Fig.~\ref{fig:one}, \rev{where we also illustrate $k-$dependence of the absolute values of the transmission $t=1/M_{22}$ and of the left ("L") and right ("R") reflection coefficients, i.e., of $r_{\rm L}=-M_{21}/M_{22}$ and $r_{\rm R}=M_{12}/M_{22}$.}
  
  \subsection{On $\PT$-symmetric potentials} 
  \label{subsec:PT}
  
 Construction of potentials with a prescribed SS, which at the same time are subject to additional constraints, like a requirement of obeying specific symmetry, becomes less straightforward. As an example, here we consider a possibility of finding a  $\PT$-symmetric potential, satisfying the condition $U(x)=U^*(-x)$ (such potentials attract particular attention in many physical applications~\cite{applic},  {especially in optics~\cite{PToptics}}). $\PT$ symmetry results in coincidence of SSs and time-reversed SSs, that is for all $|k_0|$ where the singularities exist, they exist simultaneously for $k=k_0$ and $k=-k_0$~\cite{Longhi}. 
 
  First of all we notice that one can not find a $\PT$-symmetric potential with SSs, simply by choosing a $\PT-$symmetric $z(x)$. Indeed,  boundary conditions (\ref{eq:z2}) for a symmetric function hold only if it has zero Cauchy data at $x=\pm 1$,  and therefore  Eq.~(\ref{eq:z1}) implies that $ z\equiv 0$, while we need a function $z(x)$ which does not vanish at $x\in [-1,1]$.  
  
  At the same time, for any $\PT$-symmetric potential $U(x)$, one can find a pair of two linearly independent  solutions $z(x)\in C^2([-1,1])$ and $\tz(x)\in C^2([-1,1])$   obeying the relation 
 \begin{eqnarray}
 \label{PTrelation}
 \tz(x)=z^*(-x). 
 \end{eqnarray}
 The Wronskian $W(z,\tz)$ of these solutions is a constant $c_0\neq 0$. Thus, if $z(x)$ is chosen in such a way that $z(x)\neq 0$ at $x\in[-1,1]$, one can find $\tz$ from the equation $W(z,\tz)=c_0$: 
 \begin{eqnarray}
 \label{z12_}
 \tz(x)=z(x)\int_{0}^x\frac{c_0}{z^2(\xi)}d\xi+c_1z(x),
 \end{eqnarray}
 where $c_1$ is a constant. From (\ref{PTrelation}) and (\ref{z12_}) it follows that the values of $z(x)$ at $x>0$ and $x<0$ are related to each other by
  \begin{eqnarray}
 \label{z12}
 z^*(-x)=z(x)\int_{0}^x\frac{c_0}{z^2(\xi)}d\xi+c_1z(x).
 \end{eqnarray}
 The inverse statement is also valid.
 If one can find a function $z(x)$ which satisfies Eq.~(\ref{z12}), and which is continuous and has continuous derivatives at $x=0$, one can construct a $\PT-$symmetric potential $U(x)$ for which $z(x)$ is a solution. 
 
 It is straightforward to verify that the continuity of $z$ and $z'$ at $x=0$ imply that constants $c_0$ and $c_1$ can not be arbitrary, but must be chosen as follows:
 \begin{eqnarray}
 \label{cont_cond}
 c_1=\frac{z^*(0)}{z(0)}, \quad c_0=-2\mbox{Re}\left\{z(0)[z^\prime(0)]^*\right\}\neq 0.
 \end{eqnarray}
 Now a $\PT-$symmetric potential in the interval $0\neq x\in [-1,1] $ can be obtained from (\ref{5}) where $z(x)$ satisfies (\ref{z12}) and (\ref{cont_cond}). 
 
 Note, that the above construction leads to the $\PT-$symmetric potentials with continuous real parts, but with possible jumps of  imaginary parts at $x=0$. Continuous $\PT-$symmetric potentials can be obtained similarly, but requiring additional conditions of the continuity of the second derivative $z''(x)$ at $x=0$.   
  
 Finally, to guarantee the existence  of a SS at $\pm k_0$, one has to satisfy boundary conditions (\ref{eq:z2}). Without loss of generality one can fix the amplitude of $z(x)$ at $x=1$, i.e., consider  
 \begin{eqnarray}
 \label{bound_PT_1}
 z(1)=1, \quad z^\prime(1)=ik_0,
 \end{eqnarray}
 and impose the requirement $z^\prime(-1)+ik_0z(-1)=0$. This last condition is equivalent to
 \begin{eqnarray}
 \label{consist}
   \int_{0}^{1}\frac{dx}{z^2(x)}=\frac{1}{2ik_0}+\frac{z^*(0)}{2z(0)\mbox{Re} \left\{z(0)[z^\prime(0)]^*\right\}}. 
 \end{eqnarray}
    Thus, in order to  construct a $\PT$-symmetric potential having a SS at $\pm k_0$, one has to choose any function $z(x)\in C^2([0,1])$ different from zero on the interval $x\in [0,1]$, which has a sufficient number of control parameters to satisfy (\ref{bound_PT_1}) and (\ref{consist}). Computing the values of the control parameters from these conditions, extending $z$ for negative $x$ by (\ref{z12}), and substituting the found $z(x)$ in Eq.~(\ref{5}) one finally obtains a $\PT-$symmetric potential with a SS at $\pm k_0$.  
  
\begin{figure}
	\includegraphics[width=0.49\columnwidth]{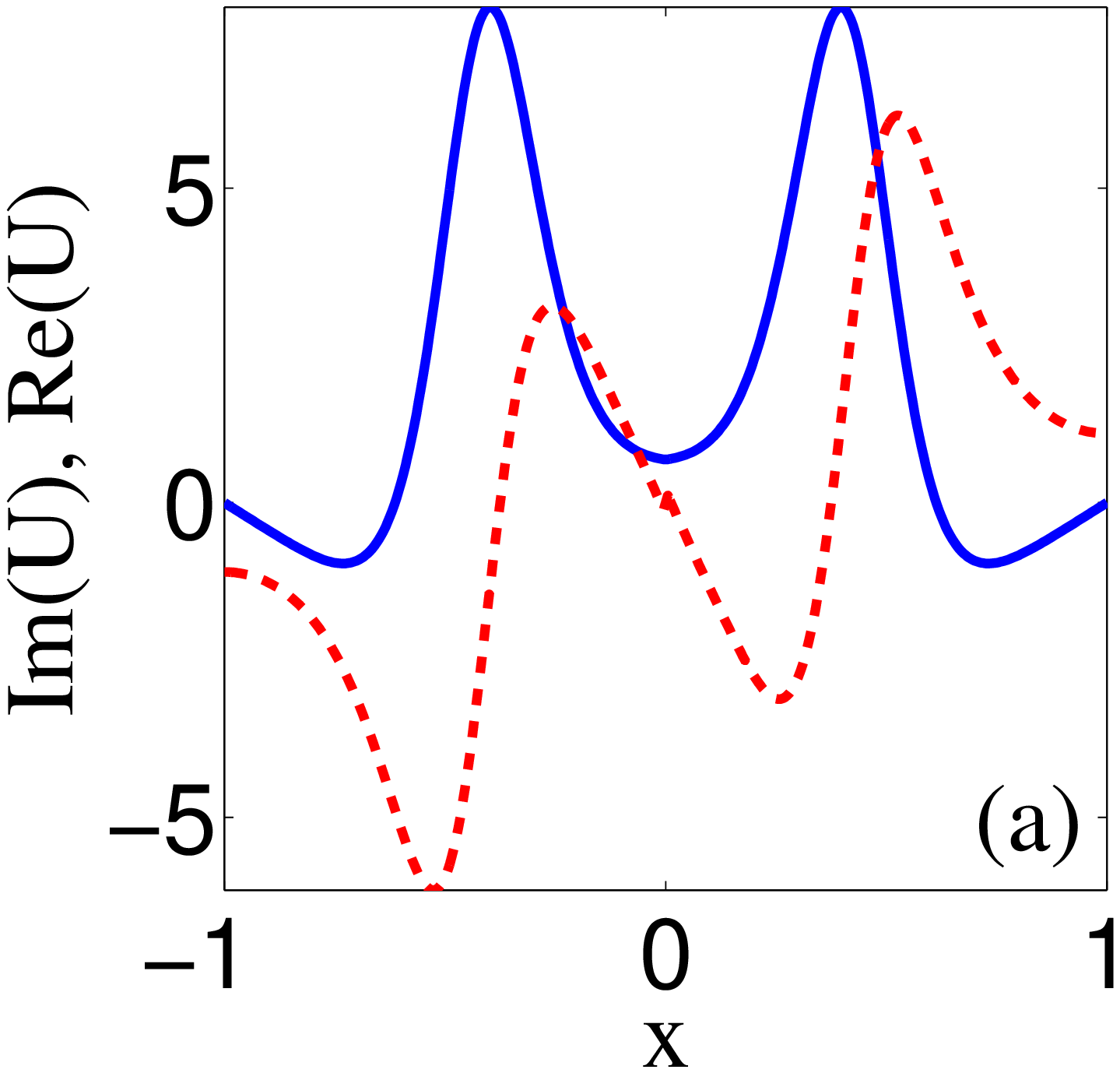}
	\includegraphics[width=0.49\columnwidth]{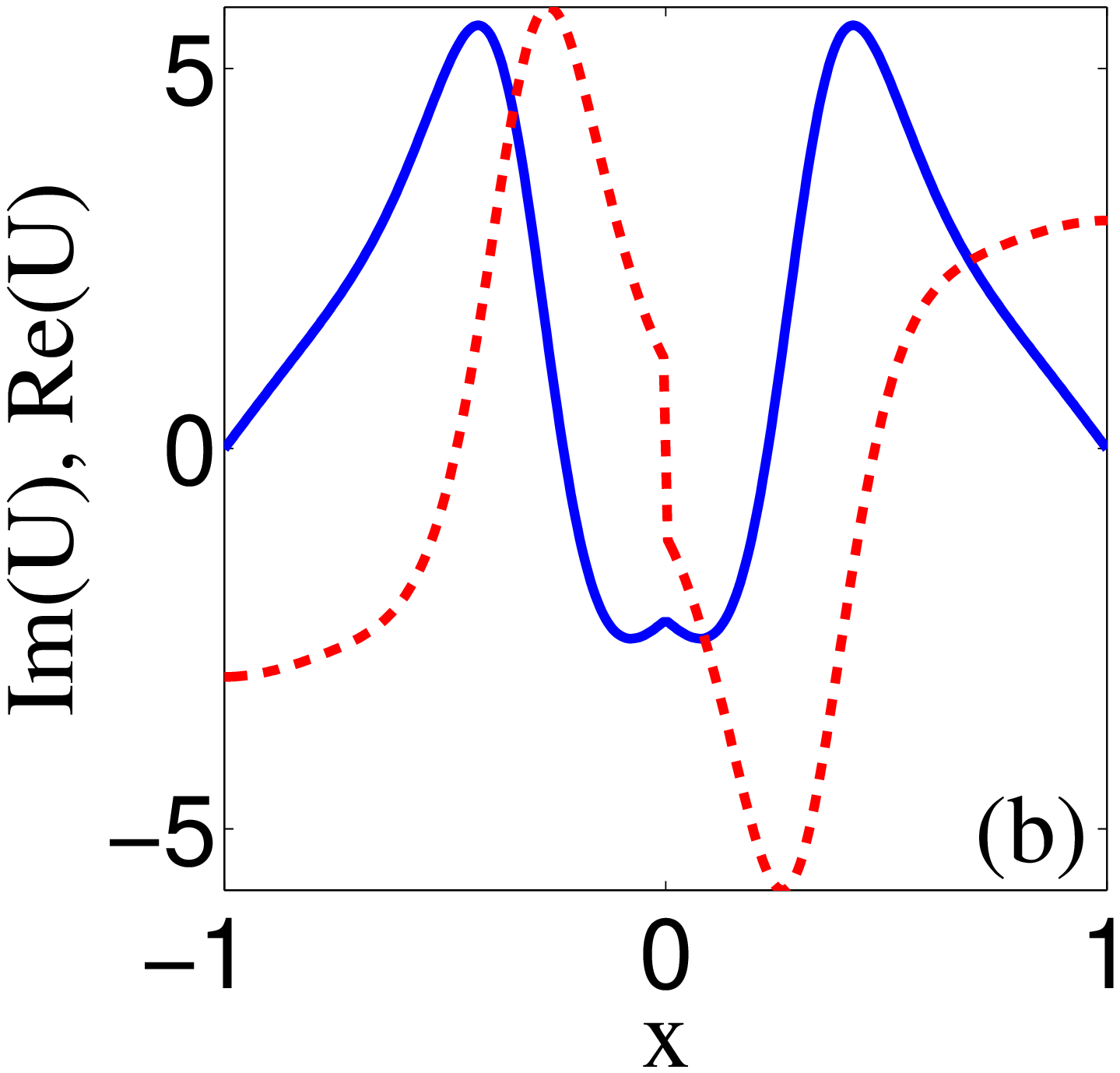}

	\caption{ Examples of the $\PT$-symmetric potentials with SSs $k_0=1.1$ (a) and  
	$k_0=3$ (b).  The real and imaginary parts of the potentials are shown by blue solid and red dashed lines, respectively.} 
			\label{fig:PT}
\end{figure}

  To illustrate this approach with an example, we consider the function
  \begin{eqnarray}
  \label{eq:ansatzPT}
  z=\left[\alpha+(\alpha-1)\cos(\pi x)+\frac{i\beta}{\pi}\sin(\pi x)  
  \right.   
  \nonumber \\   
  \left.
  +\frac{2i(k_0-\beta)}{\pi}\cos\left(\frac{\pi x}{2} \right) \right]^{-1}
  \end{eqnarray}
 with two control parameters $\alpha$ and $\beta$. In Fig.~\ref{fig:PT} we show $\PT-$symmetric potentials obtained for $k_0=1.1$ (panel a; where $\alpha  \approx -0.3073$ and $\beta\approx 0.29348$) and $k_0=3$ (panel b; where $\alpha  \approx -0.039$ and $\beta\approx4.036783$).

\subsection{On experimental feasibility of the potentials.}  
\label{subsec:experiment}

  We conclude this section with comments on experimental feasibility of the above potentials, which in similarity with previous studies~\cite{SUSY,AKSY} devoted to "engineering" potentials with prescribed properties, are theoretical constructions, so far. The step-like potentials discussed in Sec.~\ref{subsec:rectangular} (that can be straightforwardly extended to complex potentials with many steps) correspond to the most typical and widely used model of a layered optical structures containing slabs with gain and absorption.

 Potentials like the ones discussed in Secs.~\ref{subsec:varying} or \ref{subsec:PT}, have more sophisticated shapes. A suitable media for their implementation is a gas of multilevel atoms, where fairly general profiles of the dielectric permittivity can be created. For example, possibilities of generation of complex parabolic and cos-like potentials were discussed in~\cite{HHK}.  A way of creation of a complex Scarf II potential in atomic gas was described in~\cite{HGH}. For more general consideration see a recent review~\cite{HH}.

The choice of the examples in the present paper is based on simplicity of the field profiles, $\psi$ (or of the function $z(x)$) generating potentials. This  does not lead to the simplest possible shapes of the obtained potentials. If necessary, one can simplify the shape of the potential by using multi-parametric substitutions, like (\ref{u_gen}) exploited below, instead of choosing  functions $\psi$ having simple forms. Such "optimization" of potentials with SS, however, goes beyond the scope of the present work.  
 
 \rev{If the conditions required for existence of a SS are satisfied approximately, rather than exactly, the potential slightly deviates form the "exactly constructed" one. Then the zeros of $M_{22}(k)$ are also shifted from the $k_0$ value. Depending on the deformation of the potential, different scenarios can be realized. A simple or $n$-th order SS  may remain on the real axis  (see an example in the next Sec.~\ref{sec:twodelta}). A zero of $M_{22}(k)$ that defines SS may move to the lower or to the upper half parts of the complex $k-$plain.  A zero of $M_{22}(k)$ corresponding to a SS of a higher order, may split into the lower-order roots of $M_{22}(k)$.   Different aspects of dynamics and transformations of SSs under deformations of potentials were already described in the literature, see e.g.~\cite{StonePRL,Longhi,Garmon,HHK,KZ}. An example of splitting of a second-order SS is discussed below in Sec.~\ref{sec:twodelta}.   
}

 \section{Potentials with a spectral singularity of the second order.}
 \label{sec :pot_SS_second_order}

\subsection{General results}

Now we construct potentials having a SS of the second order. To this end, multiplying (\ref{1a}) by $\phi(x)$ and (\ref{v1}) by $\psi(x)$ and subtracting one equation from another we obtain an expression which can be integrated (with respect to $x$) once. Provided the solution $\psi(x){,~|x|<1,}$ is given, this results in the first order ODE for the function $\phi(x)$:
\begin{equation}
\label{vvp}
\phi'(x)\psi(x)-\psi'(x)\phi(x)+2k_0\int_{-1}^x\psi^2(\xi)d\xi=-i\alpha_{-}^2e^{2ik_0}
\end{equation}
where   we used the boundary condition (\ref{1b})   
	at $x=-1$. Inserting now $x=1$ in this formula
	and using (\ref{v1}), (\ref{v1b}), we obtain the following condition
\begin{eqnarray}
\label{doubleSScondition}
 \int_{-1}^{1}\psi^2(x)dx=\frac{\alpha_-^2+\alpha_{+}^2}{2ik_0}e^{2ik_0}, 
\end{eqnarray} 
which must be satisfied for a SS to be of the second order.

Now, Eq.~(\ref{vvp}) implies that
\begin{eqnarray}
\phi=\psi(x)\left[i+\frac{\beta_-}{\alpha_-}-i\alpha_-^2e^{2ik_0}\int_{-1}^{x}\frac{d\zeta}{\psi^2(\zeta)} \right. \nonumber \\
\left.-2k_0\int_{-1}^{x}\frac{d\zeta}{\psi^2(\zeta)}\int_{-1}^{\zeta}\psi^2(\zeta')d\zeta'\right]
\end{eqnarray}
and one  can check that this function $\phi$ satisfies (\ref{v1}), (\ref{v1b}). 
 
For example, if the function $z=\psi$ at $|x|<1$ is chosen such that  the conditions
\begin{equation}
\label{baba}
 \psi(\pm 1)=1, \,\,\, \psi'(\pm 1)=\pm ik_0, \,\,\,  \int_{-1}^{1}\psi^2(x)dx=\frac{1}{ik_0}
\end{equation}
 are satisfied, then the potential $U(x)$ obtained by Eq.~(\ref{5}) has a SS of the second order at $k=k_0$. 

There exists also an alternative approach, that does not use Theorem \ref{tt} and is based on the study of the transfer matrix: multiple SSs correspond to the points $k$ where $M_{ii}$ have zeroes of a higher order. 
Thus a SS of the second order are characterized by the two conditions, that must be satisfied simultaneously:
 \begin{eqnarray}
 \label{eq:double_SS}
  M_{22}(k)=0, \qquad \frac{dM_{22}(k)}{dk}=0 
 \end{eqnarray}
for a given positive $k_0>0$.

\subsection{Examples of potentials with spectral singularities of the second order}
\label{sec:example}

\begin{figure}
	\includegraphics[width=0.49\columnwidth]{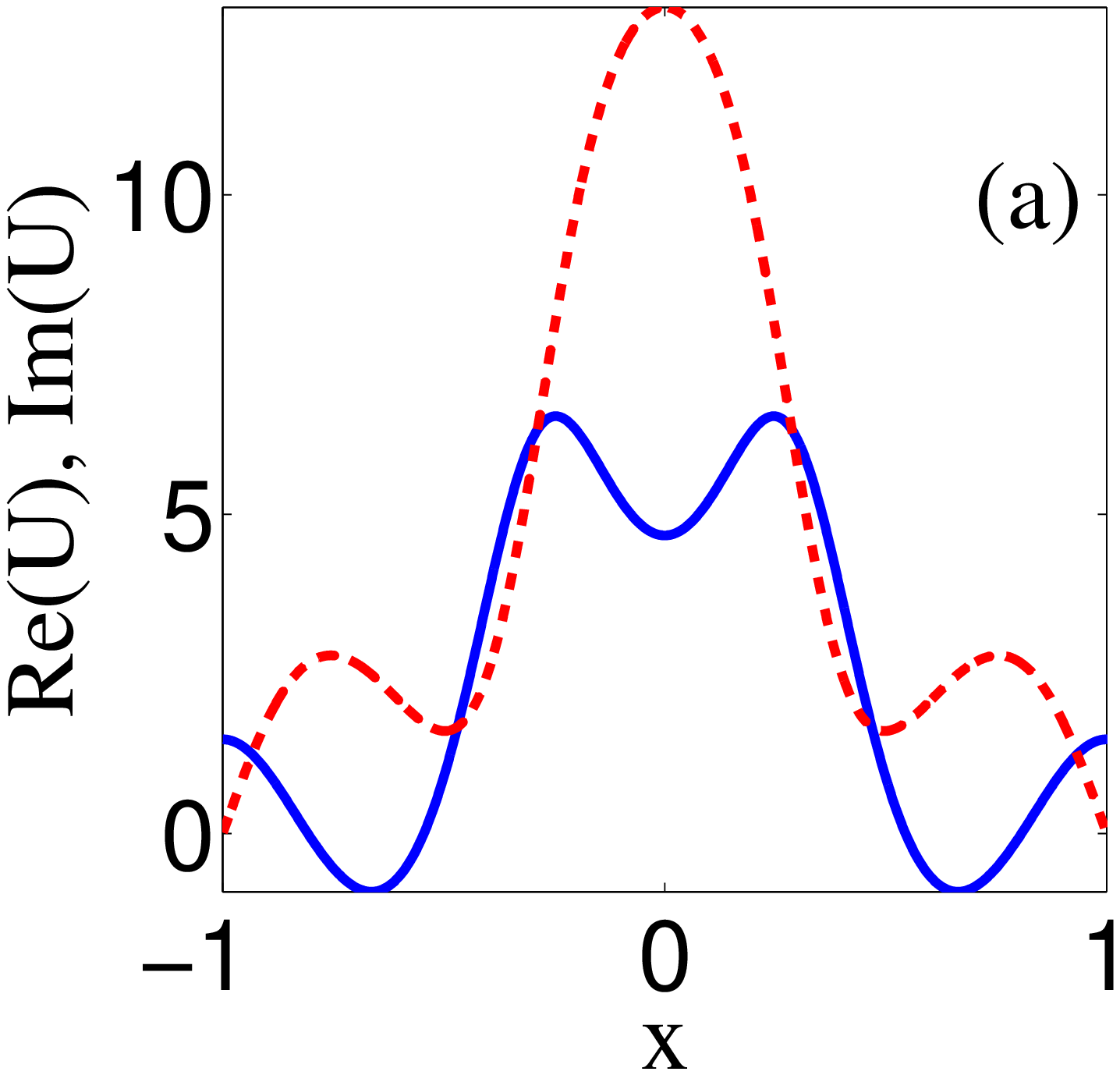}
	\includegraphics[width=0.49\columnwidth]{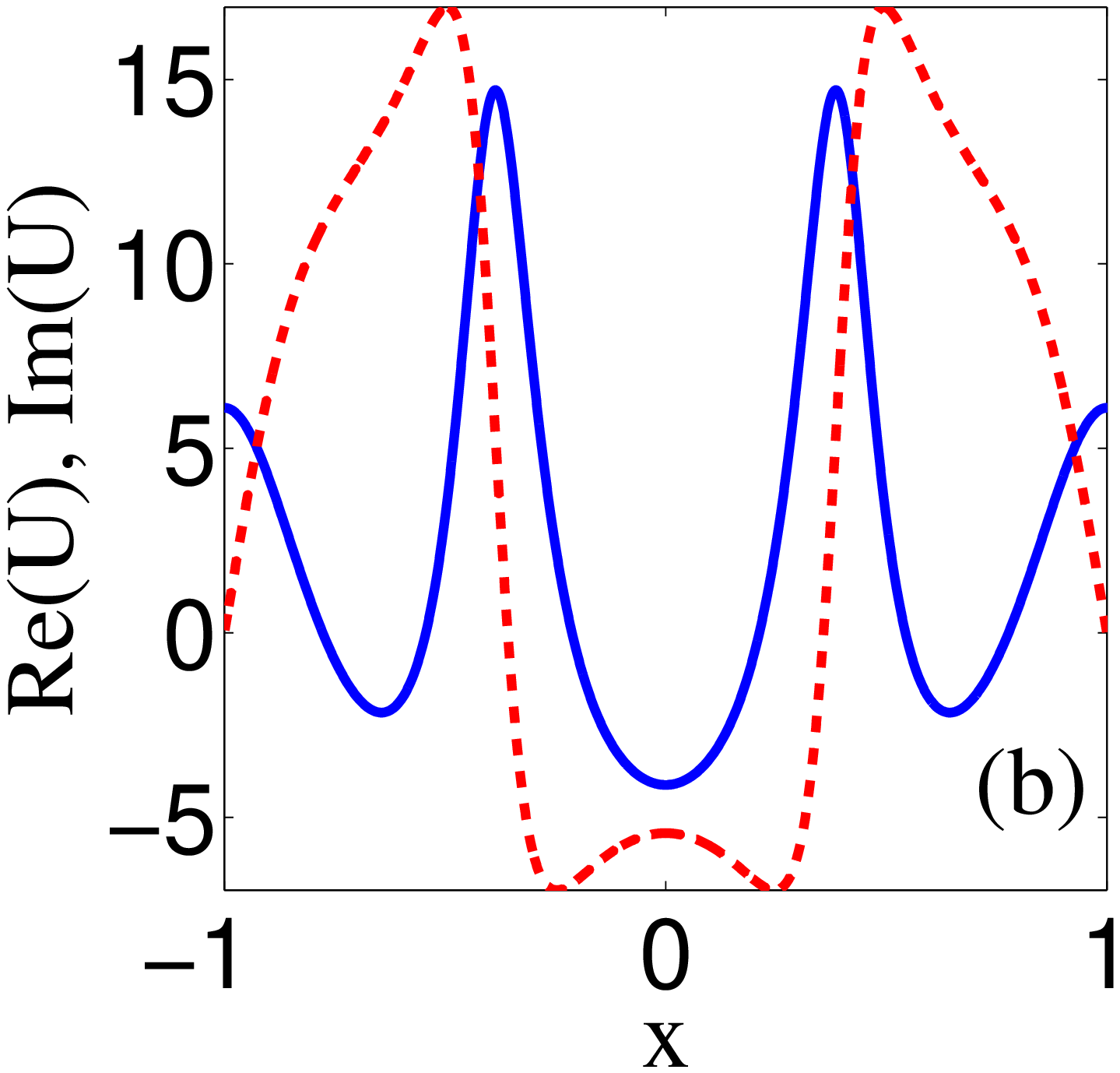}
	\caption{ Examples of non-Hermitian potentials (\ref{eq:pot_double}) having SS of the second order at $k_0=\pi$, which correspond to two different values of $a$: $a\approx -.850249$ (a) and $a\approx -.381869$ (b).  The real and imaginary parts of the potentials are shown by blue solid and red dashed lines, respectively.} 
			\label{fig:three}
\end{figure}

Multi-parametric families of potentials with SSs of the second order can be generated by {(\ref{5}) using} finite sums (or eventually convergent series) {for $z=\psi,~|x|<1,$}. Consider, for example,
\begin{equation}
\label{u_gen}
\psi=\sum_{n} \left[a_n\cos(n\pi x)+i b_n\cos\left(\frac{2n-1}{2}\pi x\right)\right] ,~|x|<1.
\end{equation}
Such ansatz requires at least four arbitrary constants $a_n$, $b_n$ to satisfy (\ref{baba}). A particular example is given by
\begin{eqnarray}
\label{eq:pot_double}
\psi= 1+a+a\cos(\pi x)+ib\cos\left(\frac{\pi}{2}x\right)
\nonumber \\
+\frac{i(\pi b+2k_0)}{3\pi}\cos\left(\frac{3\pi}{2}x\right) , ~|x|<1,
\end{eqnarray}  
where  {$a$ and $b$ are parameters to be defined. Here we already fixed two parameters ensuring that the first two relations in (\ref{baba}) are hold for arbitrary $a$ and $b$.  The respective potential has at least a simple SS at $k=k_0$. In order to obtain a second-order SS, the remaining real parameters $a$ and $b$ must be defined to  satisfy also the last condition in (\ref{baba}). This leads to the relation 
$$
b=\frac{k_0}{4\pi}\frac{5-4a}{5+8a}
$$
and to an algebraic equation of the forth order with respect to $a$. Real roots of this equation can be found numerically for a prescribed $k_0$.
In Fig.~\ref{fig:three} we show two examples of the real and imaginary parts of potentials having a SS of the second order at $k_0=3$. These potentials correspond to two different real values  $a$.}   

\section{Two Dirac-delta potentials. Splitting of a second-order spectral singularity.}
\label{sec:twodelta}

A SS of the second order, discussed in the previous section, requires specific relations among parameters of the potential.  
When these parameters are changed, the respective second-order pole of the resolvent splits in two simple ones. As discussed above, in a general situation the emergent poles do not belong to the real axis, unless the change of the parameters is properly chosen.

In this Section we discuss splitting of a SS of the second order using an example of a non-Hermitian potential in a form of two delta-functions, i.e., 
\begin{equation}\label{2d}
U(x)=\zeta_1\delta\left(x+\frac 12\right)+\zeta_2\delta\left(x-\frac 12\right), 
\end{equation}
where $\zeta_{1,2}$ are complex numbers. Our starting point is the potential $U(x)$ with a prescribed SS $k_0$.

Let $k$ be the wavenumber of the incident, reflected and transmitted waves. {Let also $M_{\zeta}$ be the transfer matrix of the delta potential with the strength $\zeta$, located at point $x=x_0$, i.e.,
\begin{eqnarray}
M_{\zeta,x_0}(k)=\frac{1}{2k}\left(
\begin{array}{cc}
2k-i\zeta  & -i\zeta e^{-2ikx_0} \\
i\zeta e^{2ikx_0}   & 2k+i\zeta
\end{array}
\right)
.
\end{eqnarray}  
Then the transfer matrix of a linear combination of delta-potentials is the product of corresponding transfer matrices. In particular, the transfer matrix of two-delta potential (\ref{2d}) is computed as 
\begin{eqnarray}
M(k)= M_{\zeta_2,-1/2}(k)M_{\zeta_1,1/2}(k).
\end{eqnarray}
Since we are interested in SSs, we analyze the diagonal element only 
\begin{equation}
M_{22}(k)
=\frac{z_1}{4k^2} \left(e^{-2ik}-1\right)+\frac{i}{2k}z_2+1. 
\end{equation}
Here for the sake of convenience we have introduced $z_1=\zeta_1\zeta_2$ and $z_2=\zeta_1+\zeta_2$ (cf. \cite{2delta,Uncu}).

  System (\ref{eq:double_SS}) consists of two linear equations for the parameters $z_{1,2}$. Its solution is
  \begin{equation}
  z_1=\frac{4k_0^2e^{2ik_0}}{1+2ik_0-e^{2ik_0}}, \quad z_2=\frac{4ik_0\left(1+ik_0-e^{2ik_0}\right)}{1+2ik_0-e^{2ik_0}}.
  \end{equation}
  This readily gives the expressions  
  \begin{equation}
     \zeta_j=\frac{z_2}{2}+(-1)^j\sqrt{\frac{z_2^2}{4}-z_1}, \qquad j=1,2,  
  \end{equation}
  for complex amplitudes   in (\ref{2d}) for which a SS of the second order exists at the prescribed wavenumber $k_0$.  
  
   We do not present here the explicit expressions of $\zeta_j(k_0)$, which are bulky. Instead, in Fig.~\ref{fig:four} (a) we plot the characteristics of the non-Hermitian two-delta potential as functions of the SS wavenumber $k_0$   (obviously there exists another pair of solutions with $\zeta_{1,2}$ swapped).  Growth of the gain in one of the delta potentials (solid red line 1) with remaining all other characteristics bounded corresponds to the asymptotics $\zeta_2\sim 2ik_0+1$, $\zeta_1\sim -e^{2ik_0}$ in the limit $k_0\to\infty$. Physically, this corresponds to wavelengths much smaller than the distance between the delta-potentials, i.e.,  to $\lambda_0/L\to 0$. Notice, that the leading order of $\zeta_2$ in this limit is the SS of a single delta-potential~\cite{DiracDelta}.
\begin{figure}
    \includegraphics[width=0.49\columnwidth]{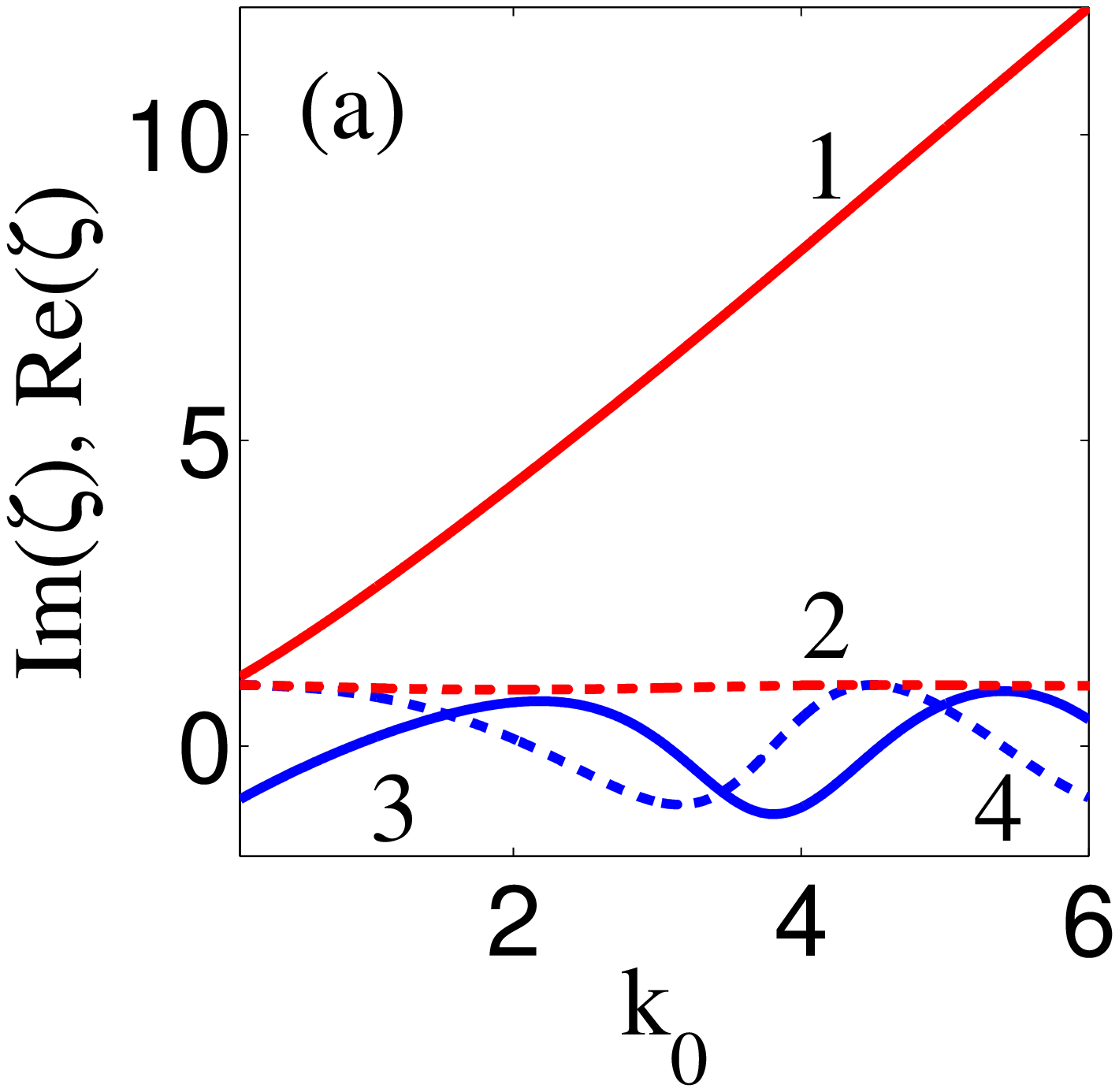}
    \includegraphics[width=0.49\columnwidth]{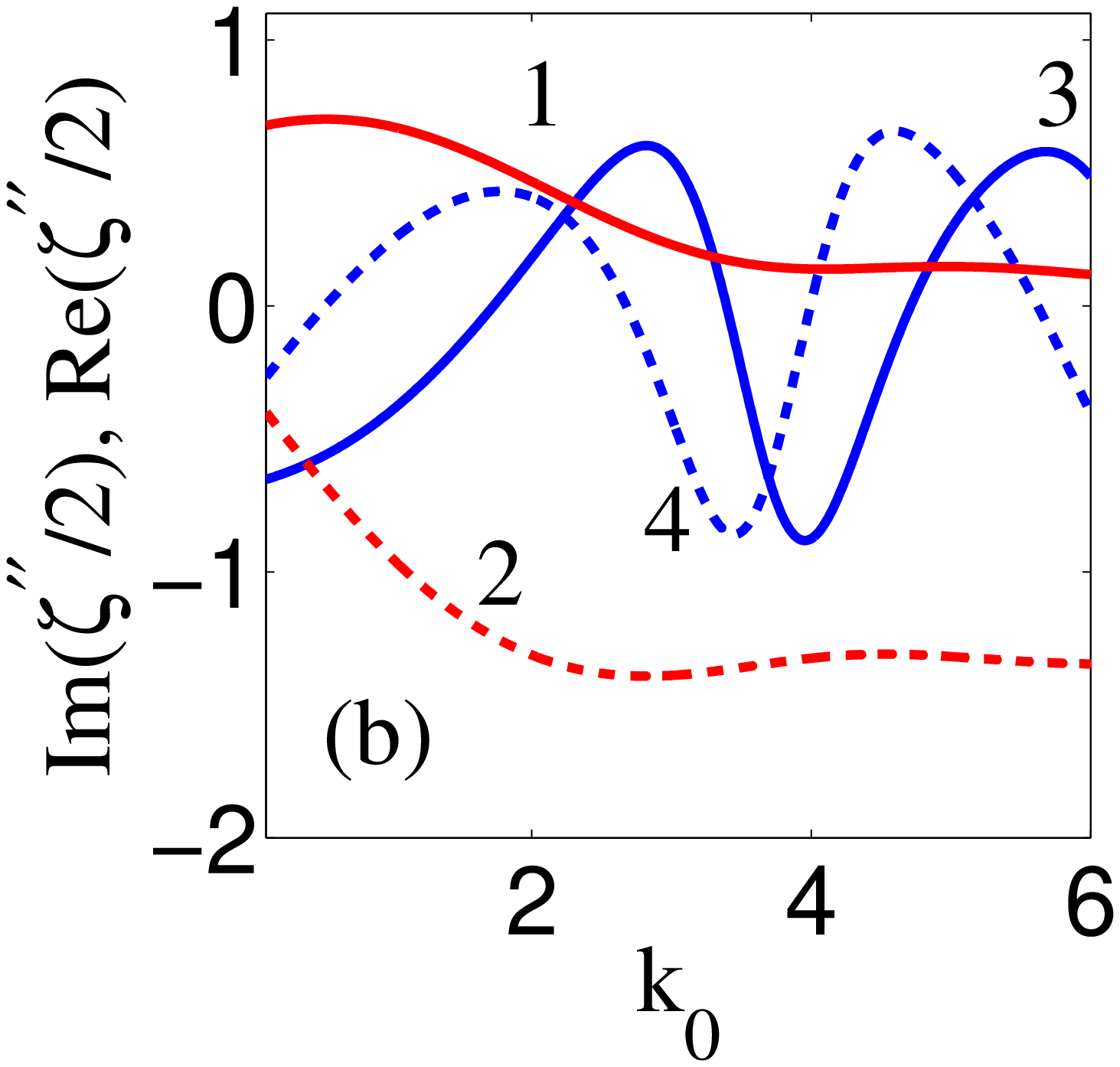}
	\caption{ Imaginary (solid lines 1 and 3) and real (dashed lines 2 and 4) parts of (a) $\zeta_{1}$ (red lines 1 and 2) and $\zeta_2$ (blue lines 3 and 4) for which there exists a positive SS of the second order with $k_0>0$, and (b) coefficients $\zeta^{\prime\prime}/2$ in Eq.~(\ref{eq:split}) describing splitting of second-order SSs.} 
	\label{fig:four}
\end{figure}

  When a potential possesses two SSs $k_1$ and $k_2$ the system can function as a CPA (if one or both $k_{1,2}$ are negative) and/or as laser (if one or both $k_{1,2}$ are positive) at the wavelengths $2\pi/k_{1,2}$. Such a possibility was considered in literature for some potentials~\cite{ScarfII-two-singul,HHK}. To find the respective parameters of the two-delta potential one has to solve the system of two equations $M_{22} (k_1)=0$ and $M_{22} (k_2)=0$ with respect to $\zeta_{1}$ and $\zeta_{2}$. 
  
  We are interested in a particular case, when the (real) roots $k_{1,2}$ emerge from splitting of a SS $k_0$ of the second order due to small variations of $\zeta_{1,2}=\zeta_{1,2}(k_0)$.
  Respectively, to find such variations, we use the ansatz $k_1=k_0-\varepsilon$, $k_2=k_0+\varepsilon$ with $\varepsilon\ll |k_0|$ and found $\zeta_{1,2}$ from the condition $M_{22}(k_0\pm\varepsilon)=0$, which guarantees the existence of SSs at $k_0\pm\varepsilon$. For small $\varepsilon$, the amplitudes $\zeta_{1,2}$, which now depend on $k_0$ and $\varepsilon$, can be found in terms of the Taylor series, in particular,
  \begin{equation}
\label{eq:split}
\zeta_j(k_0,\varepsilon)=\zeta_j(k_0)-\frac 12 \zeta_j^{\prime\prime}(k_0)\varepsilon^2+O(\varepsilon^3),~~j=1,2.
\end{equation}
 
In Fig.~\ref{fig:four} (b) we plot the real and imaginary parts of $\zeta^{\prime\prime}/2$ as functions of $k_0$ (the case $k_0>0$ is considered). We note that for splitting of a SS of the second order into two simple SSs, the required change of the imaginary parts of the complex amplitudes $\zeta_j$ (describing increase or decrease of the gain) can be of either the same sign or of different signs for two deltas, depending on the wavenumber $k_0$.
 
\rev{Splitting of the second-order SS, described by (\ref{eq:split}), resembles splitting of a simple EP under the change of the parameters of the potential. There is, however, an essential difference. After a simple EP is split it gives rise to two simple eigenvalues of the discrete spectrum. After splitting of a second-order SS the system still has singularities in the continuous spectrum; these new singularities are simple.}

The performed analysis can be generalized to  the case of $N$ delta-potentials, allowing for construction of potentials with $N$ SSs or with SSs of any order below or equal to $N$, and study the splitting of latter ones under small change of the potential parameters.

\section{Detecting spectral singularities of the second order.}
\label{sec:double_SS_detect}

Above we have constructed examples of SSs of the first and second orders. All of them correspond to zeros of the respective diagonal elements of the transfer matrix $M$. In terms of  lasing or coherent absorption of the stationary spectral problem, there is no explicit difference between those two types of SSs. This raises a question about a possibility of experimental detection of SS of different orders. In this section we show that scattering of properly designed wave-packets occurs differently allowing one to determine the type of the singularity by measuring the energy of the scattered waves.

Consider a paraxial TE beam propagating under small angle with respect to  $z$-axis.  
In the presence of an active or absorbing layer, which is  located in the interval $-1\leq x\leq 1$ (recall that we use dimensionless units) and described by a non-Hermitian optical potential $U(x)$, the wave field $\Psi$ is described by the parabolic equation
\begin{eqnarray}
\label{Schrod}
i\Psi_z=-\Psi_{xx}+U(x)\Psi
\end{eqnarray}
(coinciding with the Schr\"odinger equation, with the coordinate $z\in[0,\infty)$ playing the role of time).
One can represent solutions of (\ref{Schrod}) in the form: 
\begin{eqnarray}
\label{intervals}
\Psi(x,z)=\left\{
\begin{array}{cc}
     \Psi_{\rm int}(x,z), &  |x|\leq 1
     \\
     \Psi_\pm (x,z), & \pm x\geq 1,
\end{array}
\right.
\end{eqnarray}
 where the continuity of the field and of its derivatives must be satisfied at any $z$:
 \begin{equation}
 \label{continuity}
 \Psi_{\rm int}(\pm 1,z)= \Psi_\pm (\pm 1,z), \quad \frac{\partial \Psi_{\rm int}}{\partial x}\Bigg|_{x=\pm1 }= \frac{\partial \Psi_{\pm}}{\partial x}\Bigg|_{x=\pm 1 }  .
 \end{equation}
 
 For the sake of definiteness, we consider the wavepackets propagating away from the potential, i.e, we consider lasing.  Then the evolution of $\Psi_{\rm int}(x,z)$ is not important, and therefore we will only follow the evolution of $\Psi_\pm(x,z)$ along $z$.
We assume that the solution at $z=0$ has the form  
\begin{eqnarray}
\label{init}
\Psi(x,0)=\int_{-\infty}^{\infty} f(k) \psi_k(x) dk,
\end{eqnarray}
where $\psi_k(x)$ is a solution of (\ref{eq:1Dscat1}), (\ref{eq:1Dscat2}) for a given $k$ (here we added the subindex $k$ to emphasize this dependence), and $f(k)$ is a sufficiently well localized function assuring convergence of the integral (it describes the spectrum of the wave-packet at $z=0$). Now a solution of (\ref{Schrod}) 
is given by
\begin{eqnarray}
\label{evol}
\Psi(x,z)=\int_{-\infty}^{\infty} f(k) \psi_k(x)e^{-ik^2z} dk.
\end{eqnarray}
Hence, solution $\Psi_\pm(x,z)$ outside the interval $[-1,1]$ is determined by the integrals 
\begin{equation}
 \Psi_\pm=\int_{-\infty}^{\infty} \left[a_\pm (k)e^{ikx}+b_\pm(k)e^{-ikx}\right]f(k) e^{-ik^2z}dk,
 \end{equation}
 which must be considered for $\pm x \geq 1$.
 The continuity conditions (\ref{continuity}) are satisfied automatically.
 
 Let us consider solution (\ref{evol}) that corresponds to the wave coming from the left. Thus we assume that  $b_+(k)\equiv 0$ (i.e., there are no waves incident from the right) and $b _-(k)=1$. For such a solution, using (\ref{transfer}) we have
 \begin{equation}
  \label{psim}
  \Psi_-=\int_{-\infty}^{\infty}\!\!\left(M_{22}(k)e^{ikx}-M_{21}(k)e^{-ikx}\right)e^{-ik^2 z} f(k)dk\,\,
  \end{equation}
  and
 \begin{equation}
  \label{psip}
  \Psi_+=
  \int_{-\infty}^{\infty}\!e^{ikx-ik^2 z}f(k)  dk.
    \end{equation}
   The choice of the spectrum $f(k)$ is only constrained by the convergence of the integrals. 
   
 Let a real $k_0>0$ be a spectral singularity for a given potential, i.e., $M_{22}(k_0)=0$ (we consider a laser; the analysis of a CPA is performed similarly). When an incident wave has zero spectral width, i.e., $f(k)=\delta(k-k_0)$, one obtains the results for lasing of the monochromatic wave discussed above. Therefore now we assume that the spectral with, $\kappa$, of the incident beam is small, but nonzero, $\kappa>0$. To simplify the consideration we choose  the incident wavepacket with the Gaussian spectrum at $z=0$, i.e., 
 \begin{eqnarray}
 \label{gaussian}
 f(k)=  \frac{1}{\sqrt{2\pi}\kappa}
 e^{-(k-k_0)^2/2\kappa^2}.
 \end{eqnarray}
  The smallness of $\kappa$ is defined by the requirement for localization of the Gaussian in (\ref{gaussian}) to be strong enough for considering the elements of the transfer matrix as slowly varying functions of $k$ (see below). 
   With (\ref{gaussian}) formulas (\ref{psim}), (\ref{psip}) are rewritten as
\begin{eqnarray}
\label{psim1}
\Psi_-= \frac{1}{\sqrt{2\pi}\kappa}\int_{-\infty}^{\infty}\!\!\left[M_{22}(k)e^{ikx}-M_{21}(k)e^{-ikx}\right] 
\nonumber \\ 
\times e^{-ik^2 z} e^{-(k-k_0)^2/2\kappa^2}dk
 \end{eqnarray}
 for $x\leq -1$,  and
 \begin{eqnarray}
\label{psip1}
\Psi_+=
 \frac{1}{\sqrt{1+2iz\kappa^2}}\exp\left[\frac{i(k_0x-zk_0^2)-\kappa^2 x^2/2}{1+2iz\kappa^2}\right]
\end{eqnarray}
for $x\geq 1$. 

To interpret the obtained solution, we will use the total energy flows $I_\pm(z)$ on the left ("$-$") and on  the right ("$+$")  sides from the potential, which are defined as: 
\begin{equation}
\label{defin_I}
I_{-}(z)= \int_{-\infty}^{-1} |\Psi_-|^2dx, \quad 
I_{+}(z)= \int_{1}^{\infty} |\Psi_+|^2dx. 
\end{equation}

  First we consider the limits of this quantities at $z\to-\infty$, i.e.,  $I^{\rm in}_{\pm}= I_{\pm}(z\to -\infty)$.
  Since $\Psi_+$ is defined for $x>1$, we obtain that $|\Psi_+|^2\to 0$, and respectively $I^{\rm in}_{+}=0$. A similar result is valid for the contribution of the term containing $M_{21}$ in expression (\ref{psim1}), which is considered for $x\leq -1$. The first term in the integral in (\ref{psim1}) has the asymptotics at $z\to-\infty$, computed using the stationary phase method for a given $\kappa$ and bounded $\theta=x/(2z)$:
\begin{equation}
\Psi_-\sim\Psi^{\rm in}_{-}= \frac{1}{\kappa\sqrt{2 z}}e^{iz\theta^2 -(\theta-k_0)^2/(2\kappa^2)-i\pi/4}M_{22}(\theta).
\end{equation}
  The Gaussian in this formula describes a beam whose maximum propagates along the trajectory $z=x/(2k_0)$, i.e., the beam incident on the potential from the left. However, precisely at the maximum of the exponent one has $M_{22}(k_0)=0$. Thus for computing the total energy flow incident from the left, i.e. for estimating $I^{\rm in}_{-}$  
we have to consider the approximation 
\begin{equation}
M_{22}(\theta)=M_{22}(k_0+q )\approx \mu_{n} q^n, \quad \mu_{n} =\frac{1}{n!}\frac{d^nM_{22}(k)}{dk^n}\Bigg|_{k_0}
\end{equation}
where $n$ is the order of the SS. This approximation defines the required smallness of the spectral width: $\kappa\ll|\mu_n|/|\mu_{n+1}|$. 
Then in the limit $z\to-\infty$ one can estimate the energy flow of the incident beam as
\begin{eqnarray}
I^{\rm in}_{-}
\approx
 \Gamma\left(n+\frac 12\right)|\mu_n|^2\kappa^{2n-1}. 
\end{eqnarray}
 
Turning to the asymptotics of the output beams we consider their limits as $z\to\infty$. Obviously, $\Psi_+$ is a diffractive Gaussian beam whose maximum propagates along the direction $x=2k_0z$, i.e. away from the potential in the positive direction.  
Similarly to the above analysis, we conclude that the term with $M_{22}$ in the expression for $\Psi_-$ gives exponentially small contribution, while  the second component in (\ref{psim1}) provides the main contribution. Hence, we obtain the following asymptotics of the field $\Psi_-$ as $z\to\infty$  
\begin{equation}
\Psi_-\sim \Psi^{\rm out}_-=\frac{1}{\sqrt{2 z}\kappa}e^{-iz\theta^2-i\pi/4}e^{-(\theta+k_0)^2/(2\kappa^2)}M_{21}(-k_0).
\end{equation}
This readily yields the output energy flows  [$I^{\rm out}_{\pm}= I_{\pm}(z\to +\infty)$] 
for left ("$-$") and right ("$+$") out-propagating beams:
\begin{eqnarray}
I^{\rm out}_{-} \approx
\frac{\sqrt{\pi}}{\kappa}|M_{21}(-k_0)|^2 
, 
\qquad
I^{\rm out}_{+} \approx \frac{\sqrt{\pi}}{\kappa}.
\end{eqnarray}

 Thus, if a narrow-spectrum (small $\kappa$), weak beam with $I^{\rm in}_{-}=\mathcal{O}(\kappa^{2n-1})$ having the central wavelength $\lambda_0=2\pi/k_0$ is incident on a non-Hermitian potential,  it generates two relatively strong, $I^{\rm out}_{\pm}=\mathcal{O}(1/\kappa)$, outgoing beams provided $k_0$ is a SS of the potential. The essential difference among SSs of different orders $n$, is that in order to generate output beams with the prescribed intensities (of order $1/\kappa$ in the above example), the energy flow of the incident beam must be scaled as $\kappa^{2n}$. In particular, to initiate lasing of beams of a given intensity in the case of a potential with a SS of the second order ($n=2$), the incident beam must have $\kappa$ times smaller intensity than the intensity needed in the case of a potential with a simple SS ($n=1$).

In the case of CPA solutions (not presented here), the same phenomenon manifests itself in much smaller energy outflow, resulting from the incidence of (properly chosen) large amplitude beams on both sides of the potential. 

\section{Conclusion} 

We have formulated the approach allowing for constructing non-Hermitian potentials with prescribed spectral singularities. Focusing on one-dimensional systems we applied the method to illustrate potentials whose spectra contain one simple spectral singularity, two simple singularities, as well as singularities of the second order and their splitting. We have described an algorithm for designing $\PT-$symmetric potentials, for which the symmetry imposes additional constraints. We also discussed the effects of the different orders of spectral singularities on the scattering of wavepackets by the respective potentials.  

 {After submission of this work we get known, that the case of second order spectral singularities corresponding to a coherent perfect absorber, their splitting and physical realization (different from the considered above) was reported in~\cite{RotterStone}, where such singularities were termed coherent perfect absorber exceptional points.}  
 
\acknowledgments
 
Authors gratefully acknowledge D.A. Zezyulin for help with numerical simulations. E.L. was supported by Portuguese funds through the CIDMA - Center for Research and Development in Mathematics and Applications and the Portuguese Foundation for Science and Technology (``FCT--Fund\c{c}\~{a}o para a Ci\^{e}ncia e a Tecnologia''), within project UID/MAT/0416/2019.  B.V. was supported  by the NSF grant DMS-1714402 and the Simons Foundation grant 527180.  {We thank anonymous referee for attracting our attention to the recent work~\cite{RotterStone}.}

\end{document}